\documentclass[aps,prl,twocolumn,preprintnumbers,showpacs,nofootinbib]{revtex4}

\usepackage{graphicx}
\usepackage{dcolumn}
\usepackage{bm}
\begin{document}
\newcommand{\Od}{{\cal O}}
\newcommand{\lsim}   {\mathrel{\mathop{\kern 0pt \rlap
  {\raise.2ex\hbox{$<$}}}
  \lower.9ex\hbox{\kern-.190em $\sim$}}}
\newcommand{\gsim}   {\mathrel{\mathop{\kern 0pt \rlap
  {\raise.2ex\hbox{$>$}}}
  \lower.9ex\hbox{\kern-.190em $\sim$}}}


\title{Brane oscillations and the cosmic
coincidence problem}

\author{A.L. Maroto}
\affiliation{Departamento de  F\'{\i}sica Te\'orica,
 Universidad Complutense de
  Madrid, 28040 Madrid, Spain}%

\date{\today}

\begin{abstract}
We show that, under general assumptions,   
in six-dimensional brane-world models with
compactified large extra dimensions, the energy density of
brane oscillations scales as that of cold dark matter and its
present value is compatible with observations. Such value is 
obtained from the only dimensional scale in the theory, namely, the
fundamental scale of gravity in six dimensions $M_6\sim 1$ TeV,
without any fine-tuning or the introduction of additional mass
scales apart from the large size of the extra dimensions.
It has been suggested that the same kind of models could 
provide also the
correct magnitude of the cosmological constant. This observation
can be relevant for the resolution of the cosmic
coincidence problem in the brane-world scenario.
 \\
\end{abstract}

\pacs{95.35.+d, 11.25.-w, 11.10.Kk}
\maketitle

\section{Introduction}
Recent cosmological observations \cite{WMAP} 
seem to favour a critical or
nearly critical universe ($\Omega_{tot}\simeq 1$) with an
important fraction of cold dark matter ($\Omega_{DM}\simeq 0.23$)
and dominated by dark energy or a cosmological constant 
($\Omega_{\Lambda}\simeq 0.73$), whereas ordinary baryons only account
for a small fraction of the total energy density
($\Omega_B\simeq 0.04$). Since the time evolution
of each density parameter is  different, the fact that they
 have a comparable magnitude today suggests that, either
we are living a sort of cosmic coincidence, without any deeper explanation, 
or there is a strong relationship among the origin and evolution of the three
types of densities. 

Concerning the origin of the  dark energy, the introduction of a 
cosmological constant
could appear as the simplest {\it ad hoc} solution, 
however the required value 
$\rho_\Lambda\simeq (2 \cdot 10^{-3}$  ${\mbox{eV}})^4$
is much smaller
than any natural scale arising in particle physics models (apart from
the neutrino mass) and therefore
fine-tuning or the introduction of new physics seems unavoidable.  A more
satisfactory approach, as commented before, would be to find a link between
the evolution of dark energy and matter. This is the main idea behind the
so called quintessence models \cite{quint}, in which a light scalar field is rolling
down an exponential potential. It can be seen that  
the attractor solution of this system 
mimics the evolution of the dominant component of the universe 
(tracker solution). However, the potential parameters still require to
be fine-tuned in order to get quintessence domination today.
More recently,  tracker behaviours have also been found in the so called
k-essence models \cite{kessence} which are based on scalar fields with non-canonical
kinetic terms. They are claimed to solve the coincidence problem
without fine-tuning, (see however \cite{copeland}).

On the other hand, the most popular solutions to the 
dark matter problem (see \cite{dm} and references therein), 
i.e. the existence of an important thermal
relic abundance of weakly interacting massive
particles, such as the neutralino, or  a non-thermal background
of axions, seem to be unrelated to the cosmological constant value.
Another possible link between them was suggested in \cite{goldstone}.
There, it was shown that the existence of an ultra-light 
pseudo-Nambu-Goldstone
boson with a mass of the order
of the Hubble parameter today 
$M \sim H_0\sim 10^{-33}$ eV and a spontaneous
symmetry breaking scale $v\sim \bar M_P$, where $\bar M_P=M_P/\sqrt{8\pi}$
is the reduced Planck mass, could explain the current amount
of dark matter/energy. In fact, similarly to axionic dark matter, 
the energy density of the oscillations of the
pseudo-GB fields scales as cold dark matter, and since the initial 
amplitude of
such oscillations is set by the $v$ scale, its present density is naturally
of the correct order of mangitude $\rho_{DM}\sim M^2\bar M_P^2$
provided the oscillations have not been damped too much
by the present time. This is consistent
with the fact that,  in this scenario, oscillations
start only when $3H\lsim M$, otherwise the friction term due to the universe 
expansion
freezes the scalar field to its initial value. This condition can be written, 
using the Friedmann equation as $H^2\sim \rho_{tot}/\bar M_P^2\sim M^2$, which implies
$\rho_{tot}\sim M^2 \bar M_P^2\sim \rho_{DM}$, as commented before. 
Notice also that if today $3H \gsim M$,  the pseudo-GB energy density would 
contribute as a cosmological constant. The main
difficulty in this proposal is precisely to find a pseudo-GB candidate
whose dynamics is given by two such vastly different scales ($M$ and $v$).

More recently, the coincidence problem has been considered in the
context of large extra dimensions models. Thus for example, 
the radion field
which determines the size of the extra dimension 
has been proposed as a quintessence candidate in \cite{Albrecht}.
Also radion oscillations have been shown to potentially 
contribute to the present
acceleration of the universe or as a dark matter energy density 
\cite{peri}.
A very interesting observation was given in \cite{Sundrum,Chen}
and it is the  fact that in the Arkani-Hamed, Dimopoulos,
Dvali (ADD) brane-world model \cite{ADD}, with a fundamental
gravity scale $M_D\sim 1$ TeV,  the size of the compactified 
extra dimensions
is comparable to the dark energy scale $R_B^{-1}\sim 10^{-3}$ eV, provided
the total number of dimensions is $D=6$. 
Thus, the vacuum energy density due to loops of 
light fields propagating in the
bulk space is $\Od(R_B^{-D})$ \cite{Weinberg}. Integrating
the extra-space volume,  the corresponding contribution to the 
four-dimensional cosmological constant has precisely the correct  order
of magnitude $\rho_\Lambda\sim R_B^{-4}$. This fact has lead to the
construction of different dark-energy models with dynamical moduli fields
\cite{Pietroni, Peloso}. An additional interesting property of 
six-dimensional models is the fact that the brane tension does not
contribute to the brane cosmological constant, its only effect is
the generation of a deficit angle in the bulk metric. This observation
suggests that the amount of fine-tuning needed to solve
the cosmological constant problem could be reduced in these models
\cite{Chen,quevedo}.   

In this paper we will explore the dark matter problem in  six-dimensional
brane-world models, considering the brane as a dynamical
object which can move and fluctuate along the extra dimensions. 
The fields parametrizing the position of the brane in the extra space are
known as branons, and they have been shown to be natural candidates
to dark matter, both as thermal \cite{dark} and non-thermal 
\cite{nature} relics. In addition, such
fields can be understood as the Goldstone bosons corresponding to
the spontaneous breaking of the traslational isometries in the extra space
\cite{sun,doma}.
When such isometries are explicitly broken, due to the curvature of the
extra dimensions, such fields are no longer massless and therefore
they could play the role of the pseudo-GB fields in the 
reference \cite{goldstone}.
In this work, we will show that the mass $M$
and the spontaneous breaking scale $v$ are naturally of the correct
order of magnitude, provided the bulk vacuum energy is $\Od(R_B^{-6})$,
i.e., comparable to that generated by quantum effects in the
compact dimensions.

The paper is organized as follows. In section 2, we review the main
properties of branon fields and relate their mass to the curvature
of the bulk space. In section 3, we study the possibility
that dark matter could be in the form of brane oscillations. Section 4 is 
devoted to an explicit example based on the so called 
anti-de Sitter (AdS$_6$) soliton
solution, and finally in section 5 we discuss some difficulties
in the builiding of realistic models and 
give the main conlusions of the work. 
 
\section{Brane fluctuations}

Let us consider \cite{collider,sky} our four-dimensional space-time $M_4$ to be
embedded in a $6$-dimensional bulk space which for simplicity 
will be assumed to be of the form $M_6=M_4\times B$, where $B$ is a
given  compact manifold. The coordinates parametrizing the points
in $M_6$ will be denoted by $(x^{\mu},y^m)$, where the different
indices run as $\mu=0,1,2,3$ and $m=4,5$.

The bulk space $M_6$ is endowed with a metric tensor that we will
denote by $G_{MN}$, with signature $(+,-,-...-,-)$. For
simplicity, we will consider the following  ansatz:
\begin{eqnarray}
 G_{MN}&=&
\left(
\begin{array}{cccc}
\tilde g_{\mu\nu}(x,y)&0\\
0&- g'_{mn}(y)
\end{array}\right),
\label{metric}
\end{eqnarray}
where 
\begin{equation}
\tilde g_{\mu\nu}(x,y)=g_{\mu\nu}(x)(1+\sigma(y^2)), 
\label{massej}
\end{equation}
with $\sigma(0)=0$ and $y^2=(y^4)^2+(y^5)^2$. We have chosen 
$y^m$ as normal and geodesic 
coordinates at least with respect to 
the $y=0$ point, so that we can write $g'_{mn}=\delta_{mn}+\Od(y^2)$.
In the models in which we will be interested, the warp factor
in the metric will be
almost irrelevant since, as will we see below, $\sigma(y^2)\ll 1$.
In such a case, the usual relation of the ADD models
between the four and D-dimensional gravity scales $M_P^2=V_BM_D^{D-2}$
still holds, so that if we fix $M_D\sim 1$ TeV, the size of
the extra dimensions should be $R_B^{-1}\sim 10^{-3}$ eV.

We will work in the probe-brane approximation, so that we assume
the
3-brane is moving in the background metric given by (\ref{metric})
which is not perturbed by its presence. 
The position of the brane in the bulk can be parametrized as
$Y^M=(x^\mu, Y^m(x))$, where we have chosen the bulk coordinates
so that the first four are identified with the space-time brane
coordinates $x^\mu$. We assume for simplicity that the ground
state of the brane corresponds to $Y^m(x)=Y^m_0=0$. 

When $\sigma(y^2)\equiv 0$ we will assume that the $M_6$ isometry group
can be written as $G(M_6)=G(M_4)\times G(B)$. The presence of the brane
will break spontaneously all the $B$ isometries except for those
that leave the point $Y_0$ unchanged. In other words the group
$G(B)$ is spontaneously broken down to $H(Y_0)$, where $H(Y_0)$
denotes the isotropy group  of the point $Y_0$.
The excitations of the brane along the (broken)
Killing fields directions of $B$ correspond to the zero modes and
they are parametrized by the GB fields $\pi^\alpha(x),\; \alpha=4,5$ 
that can be
understood as coordinates on the coset manifold $K=G(B)/H(Y_0)$.
Let us assume that the number of GB fields 
equals the dimension of $B$. In that case we can choose the
coordinates on $B$ and $K$ so that
\begin{equation}
\pi^\alpha(x)=\frac{v}{R_B}\delta_m^\alpha Y^m(x)
=f^2\delta_m^\alpha Y^m(x), 
\end{equation}
where
\begin{equation}
v=f^2 R_B,  \label{rel}
\end{equation}
is the size of $K$, $R_B$ is the radius of $B$ and $f$
is the brane tension scale.

In the general case, $\sigma(y^2)\neq 0$,  the $G(B)$
isometries will be both spontaneous and explicitly broken.
Thus, expanding around $y=0$, the induced metric on the brane 
is written in
terms of branon fields as:
\begin{eqnarray}
g_{\mu\nu}(x,\pi)\;\;=\;\;\tilde g_{\mu\nu}(x,Y)
&-&\frac{1}{f^4}
\partial_{\mu}\pi^\alpha
\partial_{\nu}\pi^\alpha \\
=
g_{\mu\nu}(x)\left(1+\frac{M^2\pi^2}{4f^4}\right)
&-&\frac{1}{f^4}\partial_{\mu}\pi^\alpha
\partial_{\nu}\pi^\alpha +\Od(\pi^4)\nonumber
\end{eqnarray}
with $\pi^2=(\pi^4)^2+(\pi^5)^2$ and $M^2=4\sigma'(0)$
where the prime denotes derivative with
respect to $y^2$.

Introducing this expansion into the Nambu-Goto action for the brane
we get, up to terms quadratic in the $\pi$ fields:
\begin{eqnarray}
S_{Br} &=&-f^4\int_{M_4}d^4x\sqrt{g(x,\pi(x))}
=-\int_{M_4}d^4x\sqrt{g(x)} f^4\nonumber \\
&+&\int_{M_4}d^4x\sqrt{g(x)}\frac{1}{2}
\left(g^{\mu\nu}\partial_{\mu}\pi^\alpha\partial_{\nu}\pi^\alpha
-M^2\pi^2
\right).\nonumber \\
\label{Nambu}
\end{eqnarray}
Notice that the warp factor is responsible for the generation of
a mass term for the branon field. Thus, expressing the
bulk Ricci tensor in terms of its lower-dimensional
counterparts,  it is possible to relate the branon mass to the bulk 
curvature as:
\begin{eqnarray}
M^2=-\left.\frac{1}{2}(R'+R^m_{\;\;m})\right\vert_{y=0}, 
\end{eqnarray}
where $R'$ is the curvature scalar corresponding to the $g'_{mn}$ metric
and $R^m_{\;\;m}=-g'_{mn}R^{mn}$, with $R_{mn}$ the internal 
components of the Ricci tensor corresponding to the bulk metric.
Notice that this expression holds regardless the particular form
of the $g'_{mn}$ metric, provided the bulk metric satisfies the
conditions imposed above.

Branons also interact with the Standard Model (SM)
 particles through their energy momentum tensor.
Again the lowest order term  was obtained in \cite{doma,collider}: 
\begin{eqnarray}
{\cal L}_{Br-SM}&=& \frac{1}{8f^4}(4\partial_{\mu}\pi^\alpha
\partial_{\nu}\pi^\alpha-M^2\pi^2 g_{\mu\nu}+\dots)
T^{\mu\nu}_{SM}\label{lag}\nonumber \\
\end{eqnarray}
where the dots stand for higher order terms  in $\pi$ fields.  We see that 
the branon-SM interactions are controlled by the brane tension scale $f$. 

\section{Dark matter from brane oscillations}
In \cite{dark} it was shown that branons could
be produced by the freeze-out mechanism in an expanding universe
through
their couplings to the  SM particles in (\ref{lag}). Thus 
if branons decoupled early enough, their cosmological abundance today 
can be relevant and they could account for the dark matter
of the universe. However, apart from their thermal production,
it is also possible to produce branons non-thermally, very
much in the same way as in the misalignment mechanism for axions
 \cite{nature}. In such a case, due to the mass term
of branon fields, the energy density stored in the coherent
oscillations of the brane around the potential minimun could account
for the observed dark matter, provided the mass $M$ and
the oscillations amplitude have the correct magnitude.

 Let us briefly review that mechanism. 
If the maximum temperature reached in the 
universe was smaller than the branon freeze-out temperature 
$T_{RH}\ll T_f$, but
larger compared to the explicit symmetry breaking scale 
$T_{RH}\gg \lambda$
with $\lambda=(Mv)^{1/2}$, then brane fluctuations were initially
essentially massless and decoupled from the rest of matter fields.
In this case, there is no reason to expect that the position in the
extra dimension $Y_c$ at which the brane is created should coincide with
the minimum of the branon potential ($Y=0$). In general we expect 
$Y_c\sim \Od(R_B)$, i.e. $\pi_c\sim v$ within a region of size $H^{-1}$
\cite{axions}. The evolution of the branon
field is then simply that of a scalar field in an expanding universe. Thus, 
while $H\gg M$, the field remains frozen in its initial position
$\pi=\pi_c$. Below the temperature $T_i$ for which $3H(T_i)\simeq M$,
the branon field oscillates around the minimum.
These oscillations correspond to a zero-momentum
branon condensate, its energy density behaving like non-relativistic 
matter \cite{Turner}.

Let us define $\Gamma(T)$ as the total branon annihilation rate
 including annihilations
into 
SM particles and $4\pi\rightarrow 2\pi$ processes.  
In the case in which $H(T)>\Gamma(T)$ throughout the
history of the universe, 
the branon condensate energy
density essentially is not reduced by particle production, 
but only diluted by the Hubble 
expansion. In such a case the coherent brane oscillation can survive 
until present.
The specific conditions for this to happen were obtained
in \cite{nature} and they are summarized in the equation:
$T_i\simeq (M M_P)^{1/2}< T_{RH}< T_f$.
For light branons, a good estimation
for the freeze-out temperature was obtained in \cite{dark,cosmo}: 
$\log (T_f/\mbox{GeV})\simeq (8/7)\log (f/\mbox{GeV})-3.2$.
Notice that in the case in which we will be mainly interested, 
with only one fundamental scale, i.e. $f\sim M_D\sim $ TeV, the 
freeze-out temperature is $T_f\sim$ GeV, and therefore the previous 
condition 
is compatible with a reheating temperature above 
the nucleosynthesis temperature $T_{RH}> T_{BBN}\sim$ MeV.
On the other hand, in order for the above interval to exist, the branon
mass should satisfy $M<10^{-10}$ eV.

When these conditions are satisfied, we can
 calculate the energy density which is stored today in the form of 
brane oscillations. Assuming that $M$ does not depend on the temperature, 
it is given by \cite{axions,nature}:
\begin{eqnarray}
\Omega_{Br}h^2\simeq \frac{2.5 N\,v^2\,M\, T_0^3}{M_P\,T_i\,\rho_0}
\simeq \frac{6.5\cdot 10^{-20}N}{\mbox{GeV}^{5/2}} f^4\,R_B^2\,M^{1/2},
\nonumber \\
\end{eqnarray}
where $T_0$ and $\rho_0$ are the photon temperature and critical 
density today
respectively, and $N$ is the number of
branon fields. We see that in order for the branon condensate to be responsible
for the dark matter abundance $\Omega_{Br}h^2\simeq 0.1$, the branon mass
$M$ should be in the range $M=10^{-27}-10^{-35}$ eV, for $f=1-10$ TeV 
and $R_B^{-1}=10^{-3}$ eV. Notice also that in this range 
$v=f^2R_B\sim \bar M_P$. Since $H_0\simeq 10^{-33}$ eV, the brane could have 
started oscillating before the present time if $M>H_0$ (its energy density
scaling as dark matter) 
or  still be frozen at its initial point if $M<H_0$ (cosmological constant).  
In order to calculate $M$ we need to specify the bulk energy-momentum 
tensor.
In the following we will consider  the simplest non-trivial model 
in which the bulk space only contains a 
cosmological constant.

\section{An example: AdS$_6$ soliton}
Let us consider the solutions of Einstein equations in  six-dimensional
space-time with a (negative) cosmological constant $\Lambda_6$. 
When the
extra space has azimuthal symmetry and the metric depends only
on the radial coordinate $\rho$ with a periodic angular coordinate
$\theta\in [0,2\pi)$, a simple solution is given by \cite{leblond,cline}:
\begin{eqnarray}
ds^2=M^2(\rho)\eta_{\mu\nu}dx^\mu dx^\nu-d\rho^2-L^2(\rho)d\theta^2,
\end{eqnarray}
where
\begin{eqnarray}
M(\rho)=\cosh^{2/5}(k\rho);\;\;\;
L(\rho)=\frac{\sinh(k\rho)}{k\, \cosh^{3/5}(k\rho)}, 
\end{eqnarray}
with:
\begin{eqnarray}
k=\sqrt{-\frac{5\Lambda_6}{8M_6^4}}.
\end{eqnarray}
Notice that the metric is normalized so that for 
$k\rho\ll 1$, we recover the Minkowskian form.
Notice that we have assumed that the presence of the brane
has no effect on the bulk metric. However, even if we include the 
jump conditions at the brane position, it can be seen that the only
consequence would be the introduction of a deficit angle in
the $\theta$ coordinate, which is related
to the brane tension \cite{cline}. In addition, in order to compactify the
extra dimensions, it has been shown \cite{Chen,cline} that it is 
possible to truncate the extra space by introducing  a 4-brane
at a finite distance $\rho=R_B$ with an anisotropic energy-momentum tensor.

It is possible to write the above metric in quasi-Minkowskian 
extra-coordinates
$y^4=\rho\cos(\theta)$, $y^5=\rho\sin(\theta)$. Expanding around $y=0$
we find for the non-vanishing metric components in the new
coordinates:
\begin{eqnarray}
G_{\mu\nu}&=&\left(1+\frac{2}{5}k^2 y^2+\dots\right)\eta_{\mu\nu}\nonumber \\
G_{44}&=&-\left(1-\frac{4}{3}k^2(y^5)^2+\dots\right)\nonumber \\
G_{55}&=&-\left(1-\frac{4}{3}k^2(y^4)^2+\dots .
\right)
\label{extramet}
\end{eqnarray}
Notice also that the extra-coordinates curves  with either $y^4=0$
or $y^5=0$ are normal and geodesic with respect to the origin
and therefore define properly normalized branon fields, so that
we can obtain the corresponding mass as: 
\begin{eqnarray}
M^2=4\sigma'(0)=\frac{8k^2}{5}=-\frac{\Lambda_6}{M_6^4}. 
\end{eqnarray}
We see that the branon mass is determined by the bulk cosmological constant.
If we had considered instead a de Sitter dS$_6$ background solution,
$M^2$ would be negative and the system unstable.

In order to obtain $M$ we need to make some assumption about
the value of $\Lambda_6$. Thus, in principle, there would be
two natural scales for the bulk cosmological constant namely, the 
TeV scale or $R_B^{-1}$. Indeed,  as commented before 
if its origin is related to  
quantum loop effects in the bulk then it would be possible to set  
$\Lambda_6\sim R_B^{-6}$. In such a case the branon mass is: 
$M\sim 1/(M_6^2R_B^3)\sim 10^{-33}$ eV, and this value
 could give rise to the correct dark matter fraction
for a brane tension scale $f\sim$ TeV, as shown before.
Notice also that for this value of $M$, we have
$ky\ll 1$ and  $\sigma(y^2)\ll 1$
for $y^2\leq R_B^2$ as required. This means that it
is possible to neglect $k^4 y^4$ and higher-order terms in the branon
potential expansion coming from (\ref{extramet}).

On the other hand, if the bulk cosmological constant
is of order $\Lambda_6\sim$ TeV$^6$, the branon mass would be also
$M\sim $ TeV. In such a case, non-thermal production is not possible,
and  the $\sigma\ll 1$ condition is not satisfied for
large extra dimension, accordingly an exponential warp 
factor would be present in the bulk metric.

Notice that in the former example, starting from a fundamental scale
$M_6\sim f\sim$ TeV, as suggested by the solution of the gauge hierarchy 
problem,
the four dimensional cosmological constant $\Lambda_4\sim R_B^{-4}$ 
\cite{Chen}, and the dark matter energy densities would be
comparable and with the correct order of magnitude. Thus, in this model, 
the two problems, i.e. the gauge hierarchy and the cosmic coincidence, 
are related to a single
one, which is the existence of a large size for the 
extra dimensions.

\section{Conclusions and discussion}
In this paper we have studied the dark matter problem
in six-dimensional  models
with compactified large extra dimensions. 
In particular,
we have shown that if the brane is coherently oscillating in a background
metric with an appropriate curvature, the energy density of the 
oscillations 
could be seen from the four-dimensional point of view as cold
dark matter. In particular, if the background metric is of the  AdS$_6$
form with a cosmological constant scale set by the size of the extra
dimensions, then the dark matter fraction could be compatible with
the value favoured by observations. In such a case, the amplitude
of the oscillations $v\sim \bar M_P$ is set by the only dimensional
scales in the problem: $f\sim $ TeV and $R_B$. Notice that this possibility
is only present in six-dimensional models. For higher dimensions, the 
typical values  of $R_B^{-1}$ and $M$ would be much larger, and the 
corresponding value of $\Omega_{Br}$ much smaller.  

Although the model presented above provides the correct
estimation for the dark matter energy density
without including additional mass scales, it is in certain aspects 
very simplistic. Thus:

1) A flat Minkowskian metric has been assumed on the brane. The
possibility of finding more realistic FRW type solutions has been analysed 
in \cite{cline}, (see also \cite{gregory}).

2) We have taken the simplest example 
in which the bulk only contains a cosmological constant. However
quantum effects in the bulk would rather generate an anisotropic 
Casimir stress tensor.
This also poses the problem of the stabilization of the size
of the extra dimensions \cite{Pietroni,Peloso}.

3) In order to get the correct value for the four-dimensional
cosmological constant, we would need a mechanism to protect
it against brane or bulk  quantum  effects. Some proposals
include bulk supersymmetry or diffeomorphism invariance
\cite{Chen,quevedo,Peloso}.
 
The branon mass $M$ could receive radiative corrections from
SM matter loops on the brane through the couplings in (\ref{lag}).
However, in the limit of massless branons, the Lagrangian is
invariant under the shift symmetry $\pi(x)\rightarrow \pi(x)+ C$. This
guarantees that polinomial corrections in $\pi$ fields
are not generated. So, the only possible branon mass renormalization
would come from the $f^{-4}M^2\pi^2 T_{SM\;\mu}^{\,\mu}$ term in the 
Lagrangian. The natural cut-off scale
for SM loops in the brane would be $\Lambda \sim M_D\sim 1$ TeV. 
However, since in the example considered in the 
previous section we have $f\sim M_D$, the renormalized mass
would read:
\begin{eqnarray}
M^2_R=M^2\left(1+\frac{\alpha \Lambda^4}{f^4}\right)
\end{eqnarray}  
with $\alpha$ a numerical coefficient of order one or even smaller.
This means that the tiny branon mass is only renormalized
by an order one factor. 

Finally, one could worry about the possibility that such an
ultra-light branon field should have been already detected both, 
as a modification of the Newton law or in collider experiments. 
However as shown in \cite{kugo}, the fact that branons interact
by pairs (see (\ref{lag})) implies that their contribution to 
the gravitational potential
occurs only at one loop level and their effect is actually unobservable
for $f\sim$ TeV, even in the massless limit.
In addition, such value for $f$ also ensures that branons cannot
be detected in current high-energy collider experiments \cite{crem,collider}. 
However, for brane tensions in that range, their signals could be seen
in the next generation of linear \cite{collider} or hadronic colliders
\cite{prep}. 


 {\bf Acknowledgements:} 
I would like to thank A. Dobado and J.A.R. Cembranos for
useful comments. This work
 has been partially supported by the DGICYT (Spain) under the
 project numbers FPA 2000-0956 and BFM2002-01003.

\end{document}